\documentclass{mn2e}
\usepackage{epsfig}
\usepackage{times}
\begin{document}

\title[Synchrotron self absorption] 
{Synchrotron self absorption and the minimum energy of optically thick radio flares from stellar mass black holes}
\author[Fender \& Bright]
       {Rob Fender\thanks{email: rob.fender@physics.ox.ac.uk} and Joe Bright \\
       Astrophysics, Department of Physics, University of Oxford, Keble Road, Oxford OX1 3RH, UK\\}
\maketitle

\begin{abstract}
We consider the case of radio flares from black hole X-ray binaries in which the flare spectrum evolves from optically thick to optically thin, under the assumption that this is due to decreasing optical depth to synchrotron self absorption. We are able to place upper and lower limits on the size of the emitting region associated with a radio flare, and determine the synchrotron source magnetic field and energy as a function of size. The energy has a clear minimum which occurs close to the condition that the magnetic field derived from synchrotron self absorption equals that calculated from equipartition. This minimum energy estimate is independent of the rise time of the event, and so may be applied to any event for which the peak flux is measured and there is evidence for self-absorption. This is a much more accurate approach to minimum energy estimation than assuming expansion at close to the speed of light. We apply this method to four examples of optically thick radio flares and find that in each case either the filling factor of the synchrotron source is considerably less than unity, or the expansion speed is considerably less than the speed of light. The combination of unity filling factor and expansion speeds close to the speed of light is completely ruled out on energetic grounds for three of the four events we consider. The inferred slowed expansion is consistent with detailed modelling of such events which has been recently reported in the literature. The minimum power requirements associated with the flares are found to be $\sim 10^{36}$ erg s$^{-1}$, which are easily accomodated in the context of stellar mass black hole accretion at near-Eddington levels, when these flares typically occur. However, the true jet power could still be orders of magnitude higher.
\end{abstract}


\section{Introduction}

Accreting stellar-mass black holes in X-ray binary systems (BHXRBs) are well known to display phases of radio flaring, which are convincingly associated with the ejection of relativistic, synchrotron-emitting component (e.g. Fender, Belloni \& Gallo 2004; Tetarenko et al. 2017). Furthermore, these flare events are known to be associated in most, probably all, cases, with changes in the nature (rate, geometry, optical depth) of the accretion flow. Arguably the single most important measurement which can be made using the radio emission, particularly in cases where it is not spatially resolved, is an estimate of the kinetic energy release from the source, which results in particle acceleration and hence the synchrotron emission which we observe. Thus we can estimate the kinetic feedback associated with a given transient event or phase of accretion, a connection of very broad significance from the physics of particle acceleration to AGN feedback and the regulation of galaxy growth (e.g. McNamara \& Nulsen 2012, Hardcastle et al. 2019).

We typically make this estimate using the assumption that synchrotron-emitting components are close to equipartition, where the energies in electrons and magnetic field are comparable (and if they are not, this provides us with a lower limit to the energy). In order to apply this technique, one needs to be able to associate a given synchrotron luminosity with a given emitting volume (Burbidge 1956, Pacholczyk 1970). Once we can do this we can estimate the magnetic field, and the near-equal contributions to the internal energy of the emitting plasma from the accelerated particles and magnetic field.

In the case of spatially-resolved emission (e.g. supernova remnants, large-scale jets of AGN) the emitting volume may be directly estimated. However, the vast majority of events from BHXRBs, which includes {\em all} events for most sources, are not spatially resolved, but instead inferred by the observation of a flare in a radio lightcurve (flux monitoring). We may then estimate the size of the emitting region to be $v \Delta t$ where $\Delta t$ is the variability timescale, typically estimated from the rise time of the event, and $v$ is the expansion speed. A hard upper limit on the expansion speed is set by $v=c$ (ignoring any doppler factor associated with bulk motion). At the other extreme it is hard to imagine how a highly relativistic plasma can be restricted to expansion at very low speeds. 
Minimum energy estimates for such events are therefore typically made assuming equipartition and relativistic expansion speeds (e.g. Fender \& Mu\~noz-Darias 2016 and references therein).
A lower limit to the size of the emitting region is less frequently discussed but can be readily estimated from brightness temperature constraints. This is because brightness temperature of the emitting region increases with decreasing physical size, and a synchrotron source has a brightness temperature limit $T_B \sim 10^{12}$K above which inverse Compton cooling will rapidly reduce the temperature (Readhead 1994 and references therein).

Playing a similar role to the expansion speed is the {\em filling factor} $f$, which represents the fraction of the inferred volume which actually contains the emitting plasma. Even if a radio source is spatially resolved in radio images it is possible that $f < 1$ if the granularity of the source is below the angular resolution scale of the images. For events inferred from flare light curves it is just as uncertain.

A secondary constraint on the relationship between size and inferred magnetic field and internal energy may be calculated as follows. Many of the flare events observed from BHXRBs (and indeed from many related relativistic jet sources including binaries containing neutron stars and white dwarfs) also show an evolution from an optically thick spectrum (typical spectral index $\alpha \geq 0$ where the relationship between flux density and frequency is $F_{\nu} \propto \nu^{\alpha}$) in the rise phase (before the peak) to an optically thin ($\alpha \leq -0.5$) spectrum in the decline. Such evolution is qualitatively consistent with even the earliest models for variable radio sources (e.g. van der Laan 1966; see Tetarenko et al. 2017 for a recent successful application of this model) and implies that the peak in the light curve at a given frequency corresponds to an optical depth $\tau \sim 1$ at that frequency. The rising phase results from a increasing surface area during the optically thick phase, while the decay phase results from adiabatic expansion losses (which were of course also occuring during the rise phase). This optical depth condition provides another relation between source size and magnetic field.

Therefore we can place upper and lower limits on the size of the emitting region, and calculate the associated equipartition and synchrotron self absorption fields for the range of allowed sizes between these constraints. From these we can in turn calculate the minimum energy of the synchrotron-emitting region as a function of its size, under the assumption that the magnetic field derived from the self-absorption condition is correct. This is not an entirely novel approach: Scott \& Readhead (1977) first compared the sizes of AGN components derived from synchrotron self absorption measurements with the size corresponding to the equipartition field; Barniol Duran, Nakar \& Piran (2013) considered similar constraints in the context of gamma-ray bursts (with the associated relativistic corrections), and Zdziarski (2014) considered the case of synchrotron self absorption in a more complex jet model that also included the the contribution to the energy budget from baryons in a relativistic flow. The model developed here takes the simplest implementation of equipartition analysis but extends it to consider an uncertain size, as discussed above, and compares it directly to observations and fits of self-absorbed flare events from BHXRBs.

\section{Analysis framework}

In the following we consider only the case of a stationary (i.e. no relativistic bulk motion) expanding radio component with no significant contribution to the energy budget from baryons. In this sense this is a much more simplified approach than that developed and presented by other groups (e.g. Barniol Duran et al. 2013; Zdziarski 2014). Nevertheless, there is good reason to believe that at least for some events such simple models may provide a reasonable estimate for the physical conditions in jets from stellar mass black holes in X-ray binaries (e.g. Tetarenko et al. 2017). Our goal is to provide a simple yet more accurate estimator for the minimum energy and magnetic field of those events for which there is good evidence that the peak of the flare corresponds to the transition from optically thick to optically thin synchrotron emission. 

Our starting assumption is that a radio flare event has been observed at two frequencies, and shows evidence that the peak at each frequency is due to synchrotron self absorption. This should mean that the light curve is observed to peak at each frequency, but the lower frequency peaks after (and generally at a lower flux density) than the higher frequency. The radio spectral index $\alpha = \Delta \log(F_{\nu}) / \Delta \log(\nu)$ will evolve from optically thick ($\alpha \ga 0.5$) before the first (higher frequency) peak to optically thin ($\alpha \la -0.5$) after the second (lower frequency) peak. Since both peaks are due to synchrotron self absorption, it is possible to apply the following analysis to each peak, as discussed below. Furthermore, note that the pure synchrotron self-absorption spectral index of $\alpha = +2.5$ is rarely seen, probably due to the peak of the synchrotron spectrum already moving into the higher frequency band by the time the rise phase is observed. The top panels in Fig 1 show examples our four such dual-frequency spectrally evolving flares.

\subsection{The equipartition field}

In the following, all units are c.g.s. and are not listed explicitly, and we assume that the underlying electron distribution is a power law of the form $N(E) dE \propto E^{-p} dE$. The equipartition magnetic field $B_{eq}$ is given by 

\begin{equation}
B_{eq} = \left( \frac92 c_{12} L \right)^{2/7} R^{-6/7} 
\end{equation}

where $c_{12}$ is a pseudo-constant which encompasses the frequency range and spectral index (or, equivalently, energy range and electron energy index) of the emission and is given in full in the Appendix, $L$ is the integrated synchrotron luminosity 

\begin{equation}
L=4 \pi D^2 \int_{\nu_1}^{\nu_2} F_{\nu} d\nu \\ = 4 \pi D^2 F_{\nu_2} \nu_2^{-\alpha} \left( \frac{\nu_2^{\alpha+1} - \nu_1^{\alpha+1}}{\alpha+1} \right) 
\end{equation}

\noindent
and $R$ is the radius (for a spherical source). The distance to the source is $D$, the lower and upper observing frequencies $\nu_1$ and $\nu_2$, the flux density at a frequency $\nu$ is $F_{\nu}$ and the spectral index $\alpha$ as defined above. Note that for most observations (and certainly all four considered in detail below) $L$ calculated exactly as above is within a factor of order unity (see Appendix) of the approximation $L_{\sim} \sim 4 \pi D^2 F_{\nu} \nu$ (for either $\nu=\nu_1$ or $\nu=\nu_2$).

For a synchrotron emitting source with uniform magnetic field $B$ the energy in electrons is given by

\begin{equation}
E_e = \frac{c_{12} L}{B^{3/2}} 
\end{equation}

\noindent
and the energy in magnetic field is given by

\begin{equation}
E_B = \frac{V B^2}{8 \pi} 
\end{equation}

\noindent
and since these are the only absolutely necessary components of a synchrotron-emitting plasma, the total energy $E \geq E_e + E_B$.

For a source {\em for which the size of the emitting region is known} (i.e. for which $R$ is fixed), the minimum total energy occurs when $E_B = (3/4) E_e$. The actual energy content may be much more than this value, even if the field is close to the equipartition value if, for example, there is significant energy in protons. 

\subsection{The magnetic field from synchrotron self absorption}

In many models for radio flares, the evolution of the flare from rise through peak to decay phases corresponds to the evolution from high optical depth ($\tau \gg 1$) to low optical depth ($\tau \ll 1$) through a moment of optical depth unity near the peak (an early example is the model of van der Laan 1966). As discussed above, such an event will show a peak at each wavelength, delayed in time and lower in peak flux density as the wavelength increases.  Hence, if an observed flare event,
{\em observed at two frequencies} (see e.g. the four top panels in Fig 1) shows clear evidence for the role of synchrotron self absorption in the light curve, indicated by a switch from positive to negative spectral index past the peak, and also (but often less clear) higher frequencies peaking earlier, then we may estimate the magnetic field in the plasma at the point of peak flux. If the peak is due to synchrotron self absorption, then the field is given by:

\begin{equation}
B_{ssa} = k_1 F_{\nu}^{-2} \left( \frac{R}{D} \right)^4 \nu_{\tau=1}^5 
\end{equation}

where 

\[
k_1 = \left( \frac{\pi c_5}{c_6} \right)^2  (2c_1)^{-5} \left( \frac{e-1}{e} \right)^2 = 3.3 \times 10^{-61}
\]

where the constants $c_1$, $c_5$, $c_6$ are from Pacholczyk (1970) and a provided in the Appendix, $R$ is the radius of the source, $D$ is the distance\footnote{A note about constants: those beginning with $c$ are from Pacholcyzk (1970) and we keep their exact name from that work; the two we introduce in this paper take the form $k_{1,2}$.}. The flux density $F_{\nu}$ and the frequency $\nu_{\tau=1}$ (henceforth just $\nu$) are those corresponding to the $\tau=1$ condition. These are not quite those observed at the peak: for electron index $p=2$ the flux at the flare peak is 5\% greater than that at $\nu_{\tau=1}$, and $\nu_{\tau=1} = 0.7 \nu_{\rm peak}$ (see e.g. Pacholcyzk 1970). We make the correction for $\nu_{\tau=1}$ in our calculations, but do not make the 5\% correction in peak flux density since it is well within the range of the other cumulative uncertainties.

Note that this calculation for the unity optical depth condition requires you to make a choice of which frequency and corresponding peak flux density to use for your subsequent calculations, as the assumption is that both peaks are due to synchrotron self absorption and so either can be used. When we present the results of our calculations in Table 1, we present the results for each frequency flare for each event, demonstrating the relatively small scatter. The plots shown in Fig 1 correspond to the calculations using the higher frequency. In Fig B.1 in the Appendix we present more detailed plots showing the solutions for the higher and lower-frequency events alongside each other.

\subsection{Brightness temperature}

The brightness temperature is given by:

\begin{equation}
T_B = \frac{F_{\nu} c^2 D^2}{2 \pi k_B \nu^2 R^2} 
\end{equation}

\noindent
and should not exceed $10^{12}$ K for a synchrotron source, which is well tested and discussed in the context of extragalactic radio sources (e.g. Readhead 1994).

\subsection{Parametrizing the size}

As noted above, both the physical size of the emitting region and the filling factor play a similar role in affecting the emitting volume of the synchrotron plasma and hence the energy requirements. We can combine these two factors into a single parameter, the effective expansion speed, $\beta_e = (f^{1/3} v_{e})/c \leq 1$. If we consider that the maximum possible radius is given by $R_c = c \Delta t$, then the radius for a lower expansion speed is given by $R = \beta_e R_c$. Similarly, the emitting volume is related to the maximum volume by $V = \beta_e^3 V_c$ where 

\[
V_c = k_2 (\Delta t)^3 
\]

and

\[
k_2 = \frac{4}{3} \pi c^3 = 1.1 \times 10^{32}
\]

\noindent
The effective expansion speed corresponds precisely to the actual expansion speed in the case $f=1.0$, or corresponds to $f^{1/3}$ in the case that the expansion speed is $c$ but the filling factor $<1$. In reality it may be a combination of the two.

In this case we may consider the equipartition and synchrotron self-absorption fields, as well as the brightness temperature, in terms of their values for maximum emitting region size, scaled by $\beta_e$:

\begin{equation}
B_{ssa}  = B_c \beta_e^4  \\
\end{equation}

\noindent
where the $B_c$ is the magnetic field derived from the synchrotron self-absorption condition for $\beta_e = 1$:

\begin{equation}
B_c = k_1 c^4 F_{\nu}^{-2} \left(\frac{\Delta t}{D}\right)^4 \nu_{\tau=1}^5
\end{equation}

\noindent
The equipartition field as a function of $\beta_e$ is

\begin{equation}
B_{eq} = (B_{eq,c}) \beta_e^{-6/7} 
\end{equation}

\noindent where $B_{eq,c}$ is the equipartition field for the same condition $\beta_e =1$:

\begin{equation}
B_{eq,c} = \left( \frac92 c_{12} L \right)^{2/7} (c \Delta t)^{-6/7} 
\end{equation}

\noindent
Similarly, $T_{B,c}$ is the brightness temperature at source maximum size so that:

\begin{equation}
T_B =  T_c \beta_e^{-2}
\end{equation}

\noindent
where

\begin{equation}
T_c = \frac{F_{\nu} D^2}{2 \pi k_B \nu^2 \Delta t^2}
\end{equation}

The condition for $T_B \leq 10^{12}$ K defines a critical expansion speed $\beta_T$, which corresponds to the minimum allowed size for the emitting region, i.e. $\beta_e \geq \beta_T$. 

\begin{equation}
\beta_{T} = \left( \frac{T_{B,c}}{10^{12}} \right)^{1/2} \sim 7 \times 10^{-7} \left( \frac{F_{\nu} D^2}{ k_B \pi \Delta t^2 \nu^2} \right)^{1/2}
\end{equation}

\noindent
Therefore, we are constrained to evaluate $B_{Eq}$ and $B_{ssa}$ in the range $\beta_T \leq \beta_e \leq 1$.

\begin{figure*}
\epsfig{file=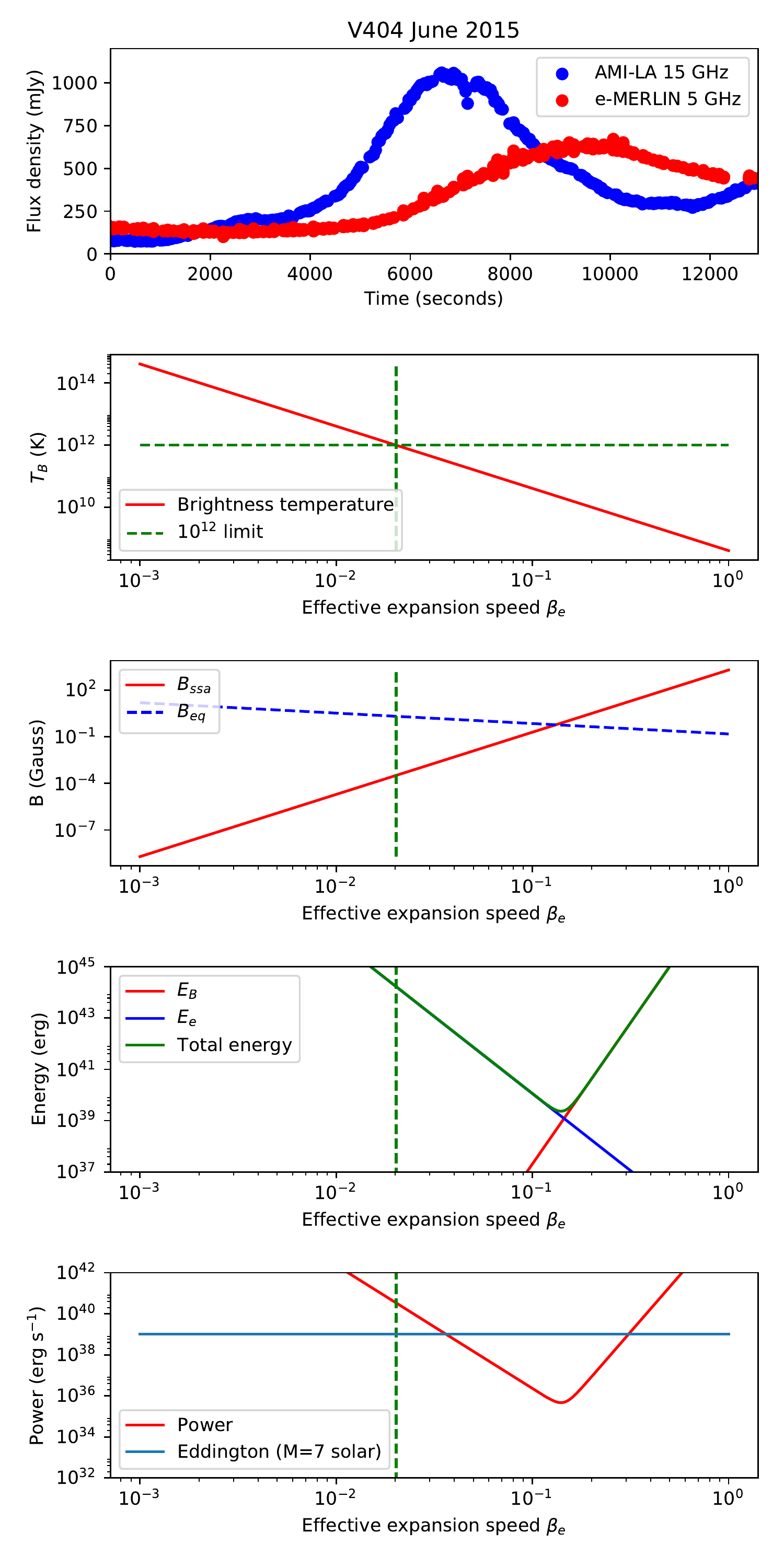, width=8cm}\quad\epsfig{file=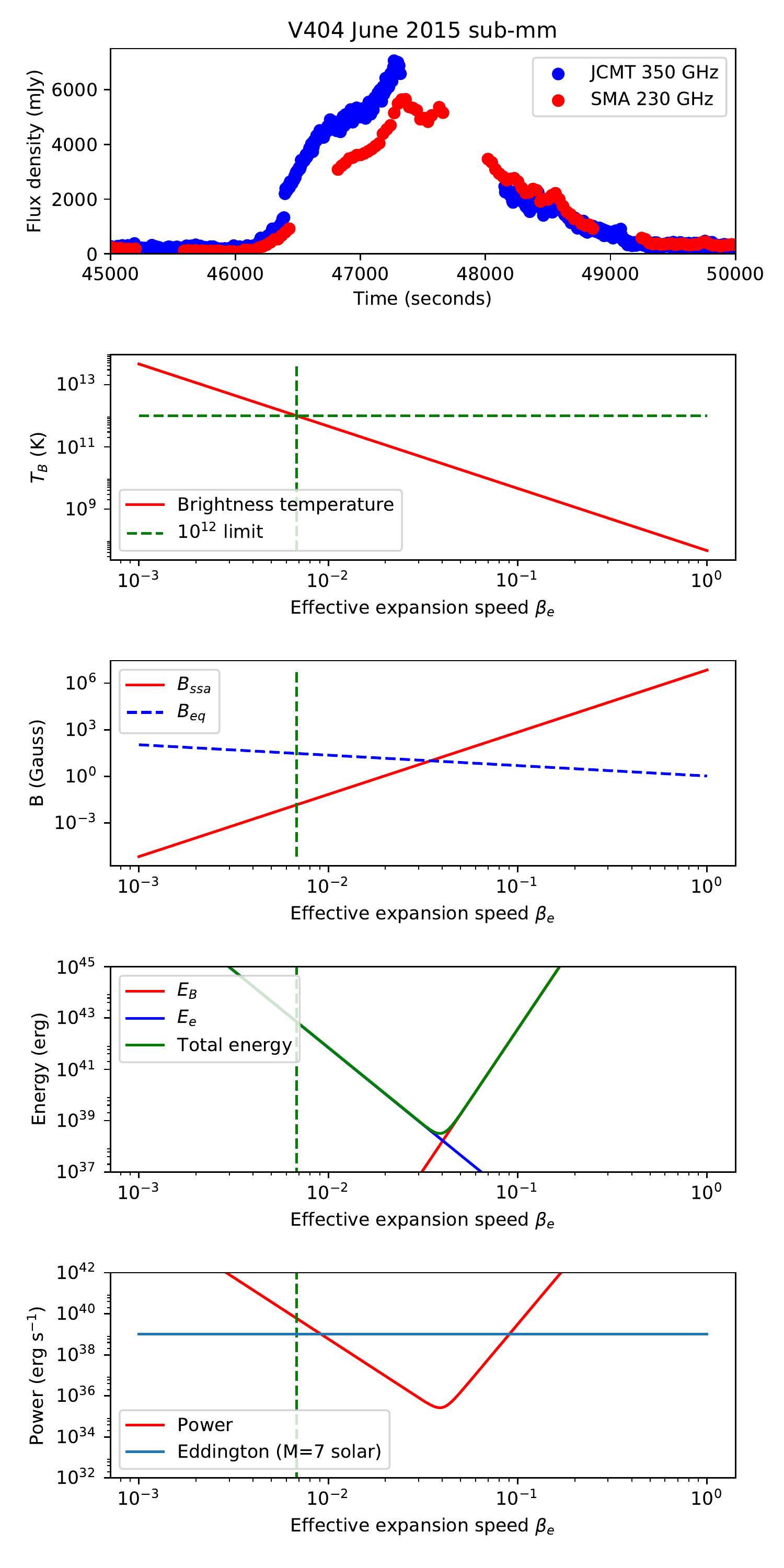, width=8cm}
\caption{Investigating the variation of magnetic field and minimum energy and power associated with an optically thick radio flare from a black hole binary. On this page both events are from the binary V404 Cyg; the left panel is a flare observed at typical GHz radio frequencies, the right panel is a flare observed at higher (sub-mm) frequencies. The top panels show the light curves at two frequencies, and demonstrate how the spectrum evolves from optically thick to optically thin through the peak, indicating that at the peak of the light curve the optical depth to synchrotron self-absorption at that frequency is $\tau \sim 1$ (see text). The second panel shows how the brightness temperature $T_B$ of the emitting region varies with effective expansion speed $\beta_e$, and how a minimum expansion speed (size) is set by the condition that $T_B \leq 10^{12}$ (vertical green dashed line). The third panel shows how the magnetic field derived from the synchrotron self-absorption condition (solid line), varies with this source size, and also how the equipartition field (dotted line) would vary with size. The fourth panel shows how the minimum energy varies with size, reaching a minimum close to the point at which the magnetic field derived from synchrotron self absorption equals the equipartition field. Finally the lowest panel indicate the required power to supply the energy on a timescale of the flare rise, and compares this to the Eddington limit for a 7 M$_{\odot}$ object (typical for stellar mass black holes).}
\end{figure*}

\setcounter{figure}{0}
\begin{figure*}
\epsfig{file=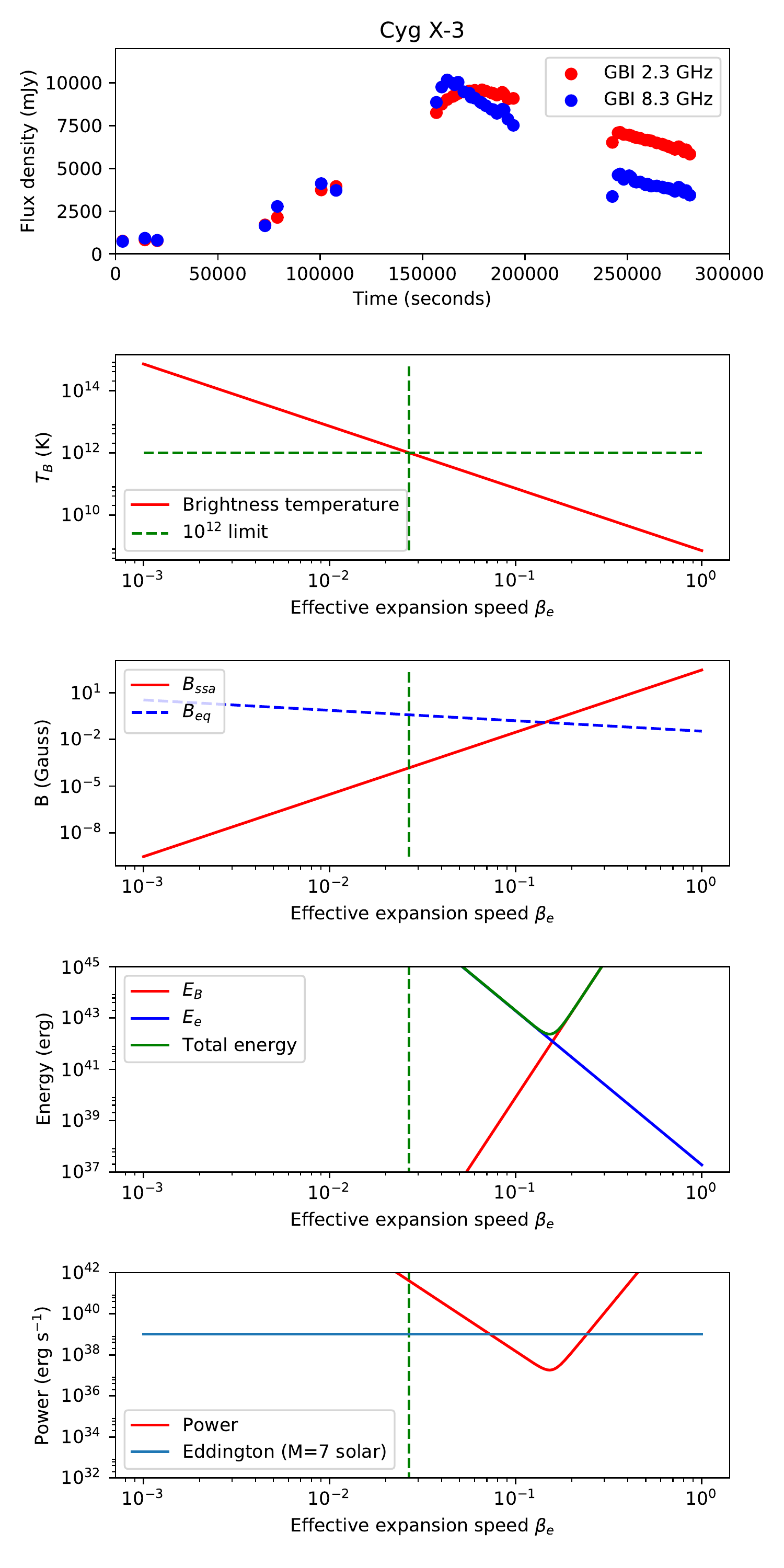, width=8cm}\quad\epsfig{file=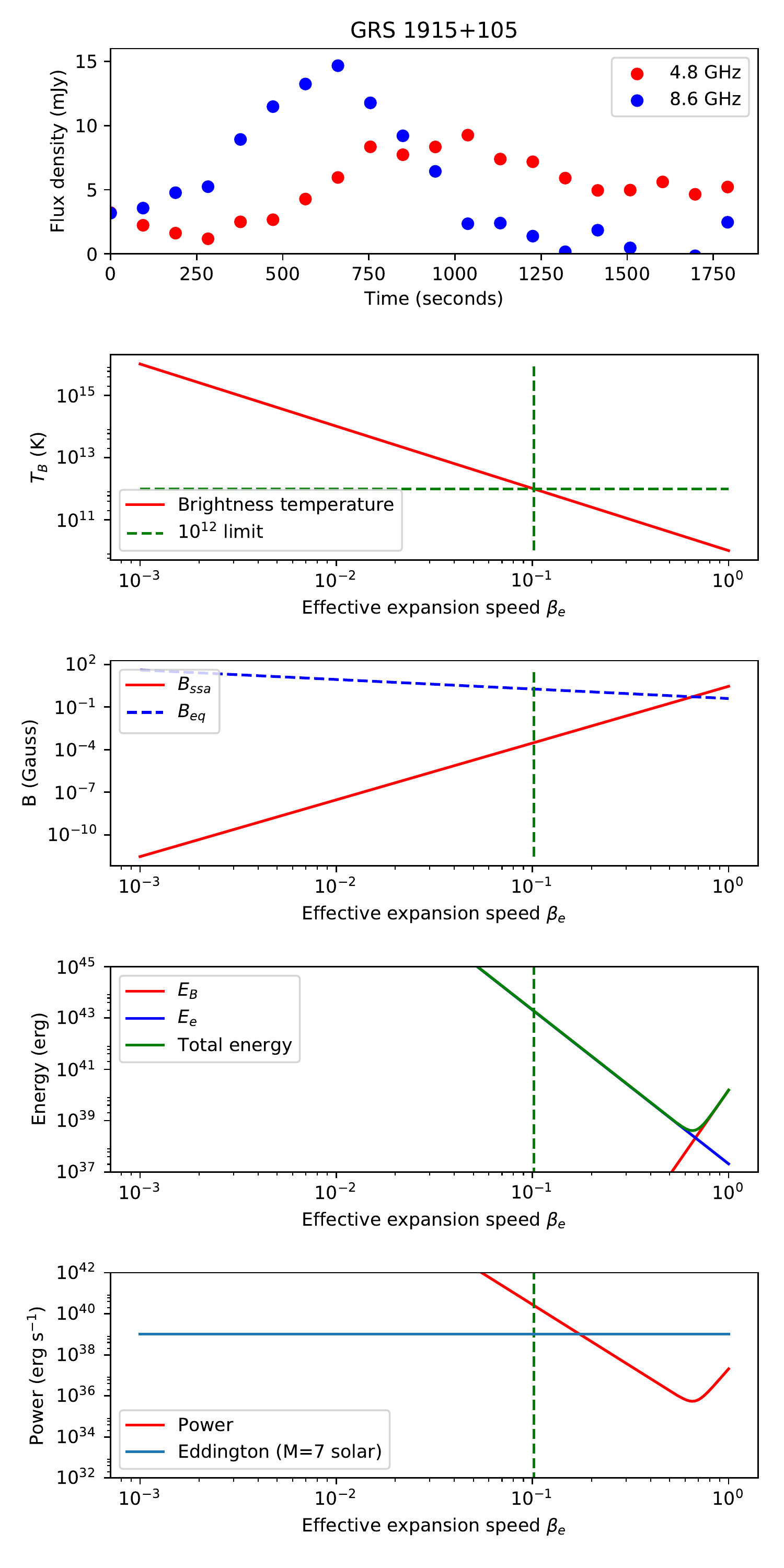, width=8cm}
\caption{{\bf{(b)}} As Fig 1(a) but for a very powerful event from the system Cyg X-3 (left panel), and the mean radio light curve of a sequence of radio `oscillations' from the system GRS 1915+105 (right panel).}
\end{figure*}

\subsection{The minimum energy condition for unknown size}

Because $B_{ssa} \propto \beta_e^4$, magnetic energy density $\propto B^2$ and volume $V \propto \beta_e^3$, the dependence of total magnetic energy content on size is extremely strong:

\begin{equation}
E_B(\beta) = \frac{V(\beta) B(\beta)^2}{8 \pi} \propto \beta_e^{11}
\end{equation}

The energy content in electrons is given by

\begin{equation}
E_e(\beta) = c_{12} B(\beta)^{-3/2} L  \propto \beta_e^{-6}
\end{equation}

Therefore there is a point at which a minimum energy is reached as a function of source size. This occurs very close to the condition where the magnetic field, as derived from synchrotron self absorption, is equal to the equipartition condition. This is equivalent to the 'equipartition radius' of Scott \& Readhead (1977). At this minimum energy as a function of size, $E_B = (6/11) E_e$ (compare to $E_B = (3/4) E_e$ for minimum energy for a fixed size source), and $E_{tot}=(17/6) E_B$. This minimum energy will occur at expansion parameter

\begin{equation}
\beta_m = \left( \frac{48 \pi c_{12} L}{11 V_c B_c^{7/2}} \right)^{\frac{1}{17}} 
\end{equation}

which simplifies to

\begin{equation}
\beta_m \simeq 1.2 \left( \frac{c_{12} L}{V_c} \right)^{1/17} B_c^{-7/34}
\end{equation}

The minimum energy, which occurs at $\beta_m$, is given by:

\begin{equation}
E_m = \frac{17 (c_{12} L)^{11/17} V_c^{6/17}}{2^{24/17} 11^{11/17} B_c^{9/34} (3 \pi)^{6/17}} \, \mathrm{erg}
\end{equation}

which simplifies to

\begin{equation}
E_m \simeq 0.6 (c_{12} L)^{11/17} V_c^{6/17} B_c^{-9/34} \, \mathrm{erg}
\end{equation}

Note that it is not clear {\em a priori} that this minimum will always lie in the allowed size range $\beta_T \leq \beta_e \leq 1$.





\begin{table*}
\scriptsize
\begin{tabular}{ccccccccccc|ccccc}\\
\hline
Flare & Source & Dist. & $\delta$ & $\nu_1$ & $\nu_2$  & $\Delta F_{\nu 1}$ & $\Delta F_{\nu 2}$ & $\Delta t_{\nu 1}$  & $\Delta t_{\nu 2}$  & $\beta_T$ & $\beta_m$ & $E_m$ & $P_m$ & $B_m$ & $T_m$\\
&& (kpc) & & (GHz) & (GHz) & (mJy) & (mJy) & (s) & (s) & & & (erg) & (erg s$^{-1}$) & (G) & (K) \\
\hline
(1) & V404 Cyg (AMI) & 2.4 & $\sim 1$ & 5.0 & 15.5 & 400 & 900 & 5000.0 & 3000.0 & 0.02 & 0.1 & $2 \times 10^{39}$ & $5 \times 10^{35}$ & 0.7 & $2 \times 10^{10}$\\
&&&&&&&&& & 0.02 & 0.1 & $8 \times 10^{38}$ & $3 \times 10^{35}$ & 0.7 & $3 \times 10^{10}$\\
(2) & V404 Cyg (JCMT) & 2.4 & $\sim 1$ & 230.0 & 350.0 & 5500 & 7000 & 1200 & 1200 & 0.007 & 0.04 & $3 \times 10^{38}$ & $3 \times 10^{35}$ & 20 & $3 \times 10^{10}$\\
&&&&&&&&& & 0.005 & 0.03 & $2 \times 10^{38}$ & $2 \times 10^{35}$ & 20 & $3 \times 10^{10}$\\
(3) & Cygnus X-3 & 8.0 & $\ga 1$? & 2.3 & 8.3 & 9000 & 9500 & 130000 & 100000 & 0.03 & 0.2 & $3 \times 10^{42}$ & $2 \times 10^{37}$ & 0.2 & $3 \times 10^{10}$\\
&&&&&&&&& & 0.01 & 0.05 & $4 \times 10^{41}$ & $4 \times 10^{36}$ & 0.5 & $3 \times 10^{10}$\\
(4) & GRS 1915+105 & 11.0 & $\la 1$? & 4.8 & 8.6 & 10 & 12 & 750 & 400 & 0.1 & 0.7 & $4 \times 10^{38}$ & $5 \times 10^{35}$ & 0.5 & $2 \times 10^{10}$\\
&&&&&&&&& & 0.1 & 0.7 & $2 \times 10^{38}$ & $5 \times 10^{35}$ & 0.8 & $3 \times 10^{10}$\\
\hline
\end{tabular}
\caption{{\em Left-most 11 columns}: Observed properties of four optically thick flares measured at upper and lower frequencies $\nu_1$ and $\nu_2$ respectively, with flare amplitudes $F_{\nu}$ and rise times $\Delta t_{\nu}$.
{\em Right-most 5 columns}: Derived parameters for four flares from BH XRBs. Note that the exact values of these estimated quantities depends upon which particular peak is being considered (see text). Hence for each event there are two rows of results, corresponding to consideration of the lower and higher frequency peaks, respectively (which is unfortunately in reverse temporal order).}
\end{table*}

The magnetic field at the minium energy condition is 

\begin{equation}
B_m = \left(\frac{48 \pi}{11} \right)^{4/17} B_c^{3/17} \left( \frac{c_{12} L}{V_c} \right)^{4/17} 
\end{equation}

which simplifies to

\begin{equation}
B_m \simeq 1.9 B_c^{3/17} \left( \frac{c_{12} L}{V_c} \right)^{4/17}
\end{equation}

Finally, the associated brightness temperature at $\beta_m$ is

\begin{equation}
T_m = \left( \frac{11}{48 \pi} \right)^{2/17} B_c^{7/17} T_c \left( \frac{V_c}{c_{12} L} \right)^{2/17}
\end{equation}

which simplifies to

\begin{equation}
T_m \simeq 0.7 B_c^{7/17} T_c \left( \frac{V_c}{c_{12} L} \right)^{2/17}
\end{equation}

\section{Relation to observed quantities}

We may recast our equations for the size and energy at the new minimum energy condition using the most easily observed quantities, namely peak flux density at some frequency, integrated luminosity, an estimate of the distance and the observed rise time. The minimum requirement for use of these expressions is, therefore, observations at two frequencies which allows the calculation of $L$ and $c_{12}$ (and which is also the minimum requirement for being certain that the flare was optically thin). We furthermore recast these observables in commonly-used units.

\begin{equation}
\beta_m = 5.5 \times 10^{-1} (c_{12} L_{\rm erg/s})^{1/17} D_{\rm kpc}^{14/17} F_{\nu, {\rm mJy}}^{7/17} \nu_{\rm GHz}^{-35/34} \Delta t_{\rm sec}^{-1}    
\label{eqstart}
\end{equation}

\begin{equation}
E_m = 1.3 \times 10^{13} (c_{12} L_{\rm erg/s})^{11/17} D_{\rm kpc}^{18/17} F_{\nu, {\rm mJy}}^{9/17}  \nu_{\rm GHz}^{-45/34}
\end{equation}

\begin{equation}
B_m = 2.4 \times 10^{-9} (c_{12} L_{\rm erg/s})^{4/17} D_{\rm kpc}^{-12/17} F_{\nu, {\rm mJy}}^{-6/17}  \nu_{\rm GHz}^{15/17}
\label{eqend}
\end{equation}

\begin{equation}
T_m = 3.6 \times 10^{14} (c_{12} L_{\rm erg/s})^{-2/17} D_{\rm kpc}^{6/17} F_{\nu, {\rm mJy}}^{3/17}  \nu_{\rm GHz}^{1/17}
\end{equation}

Of course we are really finding a minimum {\em size} when applying this approach, but when considering the physics of the jet launching and evolution, the expansion speed may be considered to be a more interesting parameter. If a radius is required, the substitution $R = \beta_m c \Delta t$ should be made.
As a result, of the four quantities at the minimum energy condition, only $\beta_m$ depends upon the observed variability timescale of the event (in such a way that substituting $R$ cancels this out). This means that the minimum energy associated with an event, as well as the associated magnetic field and brightness temperature, may be estimated simply by observing the peak flux of an event. This further implies that if a sequence of radio flares can be observed, ideally with few or not gaps in coverage, then a very reliable estimate of the minium time-averaged jet power can be made.

\section{Application to data}

We may now apply these formulae to some observed flares from black hole binaries with well-estimated distances and see how they constrain our magnetic field and energy estimates. Table 1 lists four such flare events, with the associated derived parameters; these are in turn plotted in Fig 1. Note that it is vital that the flares analysed in this way show spectral evidence for synchrotron self-absorption: we have found that there is a surprisingly large number of flare events from stellar-mass black holes which are optically thin also during the rise phase (e.g. flare V in Fender et al. 1997; Bright \& Fender {\em in prep}), in which case the peak would correspond approximately to the end of a phase of particle acceleration, and quite different physical conditions would apply.

\subsection{V404 Cyg (5--15 GHz)}
The nearby black hole binary V404 Cyg underwent a dramatic two-week outburst in June 2015. Many flares were observed simultaneously with AMI-LA at 15.5 GHz and eMERLIN at 5 GHz (Fender et al. {\em in prep}). The flare chosen is one of the clearest examples of what appears to be emission dominated by a single event, with a profile qualitatively as expected from a van der Laan (1966) model of an expanding blob. V404 Cyg is a particularly good source for this study as it has a well determined distance from radio parallax of 2.4 kpc (Miller-Jones et al. 2009) and the highest-resolution radio observations of the June 2015 outburst (Miller-Jones et al. 2019) indicate that the bulk motions of ejected components are at most mildly relativistic ($\Gamma < 1.5$).

\subsection{V404 Cyg (sub-millimetre)}
During the 2015 outburst V404 Cyg was also observed for a relatively short period at high cadence and broad frequency coverage, including up to sub-millimetre wavelengths (Tetarenko et al. 2017). This data set probes flares observed at much higher frequencies than usual. This is very important since, for optically thick flares, events observed at lower frequencies (such as the other three events discussed here) are highly smoothed out and often blended compared to what is observed at shorter wavelengths. Nevertheless, sub-mm observations such as these are less common, so we need to be able to interpret longer wavelength data; hence the comparison is useful.

\subsection{Cygnus X-3}
This is a much more distant source which frequently produces very luminous and long-duration radio flares. The data used here are from the Green Bank Interferometer (GBI) monitoring programme, which operated at 2.3 and 8.3 GHz (see e.g. Waltman et al. 1995 and Fender et al. 1997 for a discussion of Cyg X-3's behaviour as observed by the GBI). The event clearly peaks earlier at the higher frequency. There are some suggestions that the jets in Cyg X-3 are pointed towards us and the source may therefore be Doppler boosted.

\subsection{GRS 1915+105}
This is another powerful repeating jet source, albeit one with many characteristics which are different from Cyg X-3. The event analysed here is the average folded radio flare observed during a period when the source was producing optically thick radio 'oscillation' with a period of $\sim 1900$ seconds (Fender et al. 2002). Again there is clear evidence for behaviour which is qualitatively described by models such as van der Laan (1966).

\subsection{Comparison of the events}

The results for the four sources above are very interesting in the context of previous estimates of the power in jets from black hole X-ray binaries. The constraint on the expansion speed is very strong: for three of our four events (and hence two of our three sources) expansion at $\sim c$ (for a filling factor $f \sim 1$) is completely ruled out as the required power would exceed the Eddington limit by many orders of magnitude (lower panels in Fig 1). Only for GRS 1915+105 is this not so clear, and we note that this may well be the most relativistic of the events. The inferred minimum powers are well below the Eddington limit for a stellar-mass black hole, but we do remind the reader that these are very much lower limits to the jet power. The inferred magnetic fields are similar to previous estimates in the literature. 

We note that there are several other flares in the literature for which there is evidence for synchrotron self absorption. A good example is that reported by Chandra \& Kanekar (2017), in which they observed a flare from, again, V404 Cyg, which showed strong evidence for a peak due to synchrotron self absorption. They fitted this peak to be of flux density 1009 mJy at a frequency of 1.8 GHz, and estimated a rise time of $\sim 1$ day (we are ignoring the reported measurement and fitting errors, since they will be smaller than the systematics associated with the model; see Appendix). They analysed the event using the approach of Barniol Duran, Nakar \& Piran (2013) in the non-relativistic regime. As noted earlier, this is qualitatively very similar to our approach. It is reassuring therefore that the results obtained are similar: Chandra \& Kanekar (2017) calculate a minimum energy of $E_{\rm min} = 1.7 \times 10^{39}$ erg s$^{-1}$, corresponding magnetic field $B_{\rm min} = 0.25$ G and radius $4 \times 10^{13}$ cm. Our approach gives corresponding values of $E_{\rm min} = 1.1 \times 10^{39}$ erg s$^{-1}$, $B_{\rm min} = 0.08$ G and radius $7 \times 10^{13}$ cm. Using their estimate of a one-day rise time, this corresponds to an effective expansion speed at minimum energy of $\beta_m = 0.03$, very similar to the other measurements.

\section{Single frequency estimates}

We may take our approximations for the minimum energy in a flare event and reduce them to a form which is applicable to events observed at just a single frequency, under the critical assumption that they are optically thick events. We caution here that the number of optically thin flares from X-ray binaries is surprisingly large, and that, of course, without multiple frequencies it is impossible to tell (Bright \& Fender, {\em in prep}). This caution notwithstanding, we may take equations \ref{eqstart} -- \ref{eqend} and recast them substituting $L \simeq 4 \pi D^2 \nu F_{\nu}$. We furthermore need to approximate a value for $c_{12}$; looking at the calculated values for our four studied events, we choose $c_{12} \sim 10^7$ as a reasonable approximation, with an error of order unity. The formula provided in the Appendix in any case allows $c_{12}$ to be calculated for any set of assumptions about the frequency range. We may now write our single-frequency approximations as:

\begin{equation}
\beta_m = 5.6 \times 10^{1} D_{\rm kpc}^{16/17} F_{\nu, {\rm mJy}}^{8/17} \nu_{\rm GHz}^{-33/34} \Delta t_{\rm sec}^{-1}    
\label{eq1start}
\end{equation}

\begin{equation}
E_m = 1.5 \times 10^{35} D_{\rm kpc}^{40/17} F_{\nu, {\rm mJy}}^{20/17}  \nu_{\rm GHz}^{-23/34}
\end{equation}

\begin{equation}
B_m = 2.5 \times 10^{-1} D_{\rm kpc}^{-4/17} F_{\nu, {\rm mJy}}^{-2/17}  \nu_{\rm GHz}^{19/17}
\end{equation}

\begin{equation}
T_m = 3.5 \times 10^{10} D_{\rm kpc}^{2/17} F_{\nu, {\rm mJy}}^{1/17}  \nu_{\rm GHz}^{-1/17}
\label{eq1end}
\end{equation}

The very weak observable dependencies of the $T_m$, the brightness temperature at the minimum energy condition, means that this is likely to always be in the range $10^{10}$ -- $10^{11}$ K.

\section{Relativistic bulk motion}

We have only considered in this work the case where there is no relativistic bulk motion. While this is certainly {\em not} the case for some BHXRB ejection events, the bulk velocities are so poorly constrained in most cases that it is hard to accurately consider it (e.g. Miller-Jones, Fender \& Nakar 2006). Furthermore, at least in the case of the ejecta from V404 Cyg, we know that the bulk velocities are at most only mildly relativistic (Miller-Jones et al. 2019). Nevertheless, we may consider some approximate cases. The relativistic Doppler factor is given by

\[
\delta_{\rm app, rec} = \Gamma^{-1} (1 \mp \beta \cos \theta)^{-1}
\]

\noindent
where $\Gamma$ is the Lorentz factor of the bulk motion of the emitting region, $\beta$ is the bulk speed as a fraction of the speed of light and $\theta$ is the angle of the jet to the line of sight.

For discrete ejected components, the ratio of observed to intrinsic flux should vary as $F_{\rm obs} = \delta^{3-\alpha} F_0$, the ratio of observed to intrinsic emission frequencies as $\nu_{\rm obs}= \delta \nu_0$, and the observed timescale of an event as $\Delta t_{\rm obs} = \delta^{-1} \Delta t_0$. We will assume the spectral index $\alpha = 0$ at the time of peak flux. Substituting these into equations \ref{eq1start} -- \ref{eq1end}, we derive that the estimated quantities depend on the Doppler factor as

\[
\beta_{m {\rm (estimated)}} = \beta_{m {\rm (rest \, frame)}} \delta^{49/34}
\]

\[
E_{m {\rm (estimated)}}  = E_{m {\rm (rest \, frame)}} \delta^{97/34}
\]

\[
B_{m {\rm (estimated)}} = B_{m {\rm (rest \, frame)}} \delta^{13/17}
\]

\[
T_{m {\rm (estimated)}} = T_{m {\rm (rest \, frame)}} \delta^{2/17}
\]

\begin{figure}
\epsfig{file=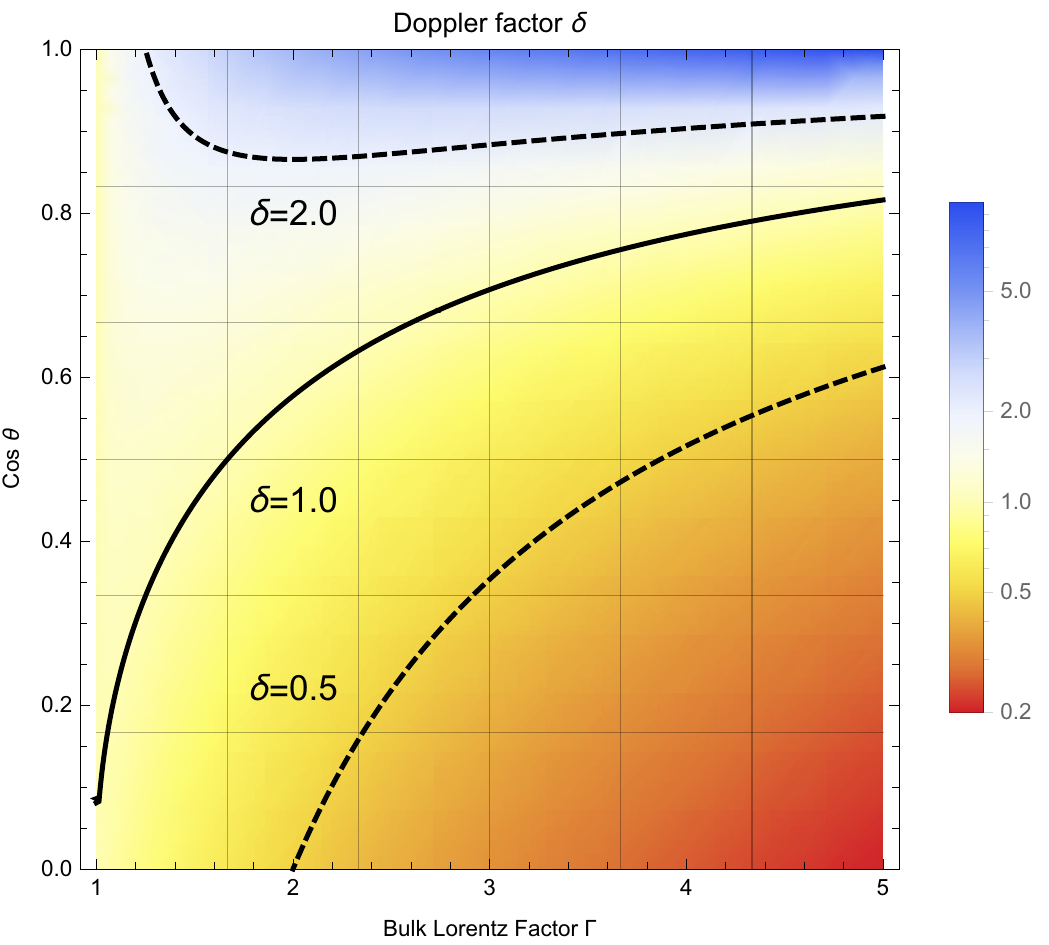, width=8cm}
\caption{Relativistic Doppler factor for the approaching component of a jet at an angle $\theta$ to the line of sight with bulk Lorentz factor $\Gamma$. We assume a uniform distribution $0 \leq \cos \theta \leq 1$ and $1 \leq \Gamma \leq 5$. Only in the region above the solid line is the Doppler factor $\delta$ greater than one (i.e. the source is boosted).}
\label{dopp}
\end{figure}

Therefore we see that all the estimated quantities increase with a positive Doppler factor. In the study of the 1997 ejecta from GRS 1915+105, Fender et al. (1999) estimated relativistic dopper factors $\delta$ for the approaching and receding jet components of 0.34 and 0.14, respectively, assuming $\Gamma = 5$. In fact under most combinations of $\Gamma$ and inclination angle, for significantly relativistic bulk motions $\delta < 1$ (see Fig \ref{dopp}). Furthermore, there will be an additional component to the minimum energy, corresponding to the bulk relativistic motion $E_{\rm bulk} \simeq (\Gamma-1) E_{\rm rest}$, so that the ratio of rest frame to estimated minimum energies will be:

\begin{equation}
R_E = E_{m {\rm(rest frame)}} / E_{m {\rm (estimated)}} \simeq \Gamma \delta^{-97/34} 
\label{erest}
\end{equation} 

\begin{figure}
\epsfig{file=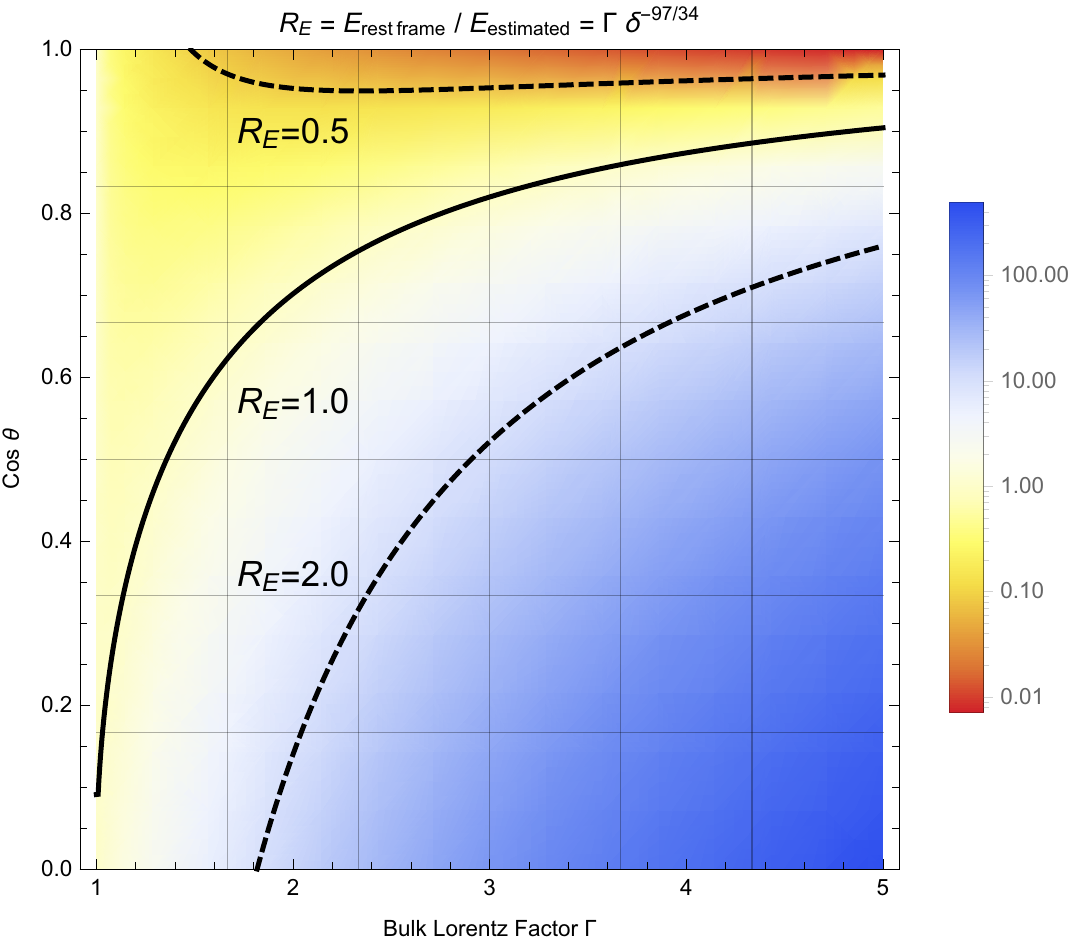, width=8cm}
\caption{Ratio $R_E$ of rest frame minimum energy, including an approximate adjustment for the bulk relativistic motion, to that estimated from observables. For the large majority of parameter space the minimum energy estimated from observables remains a genuine lower limit to the energy associated with the event.}
\label{eratio}
\end{figure}

\noindent
We plot this quantity in Fig \ref{eratio}. There is very little room in parameter space in which our {\em minimum} energy estimate will actually be a significant {\em overestimate} due to bulk relativistic motion. This is because the larger boosted fluxes tend to raise our energy (and other parameter) estimates in the same direction as the intrinsic energy increase due to the bulk motion. It is much more likely that we have significantly underestimated the minimum energy (due to $\delta < 1$, and so reduced apparent fluxes plus added bulk relativistic motion), in which case it remains a genuine lower limit.

\section{Discussion and Conclusions}

It is commonly assumed that the relativistic plasma associated with synchrotron flaring events must be expanding relativistically at speeds close to $c$, with a mean expansion speed of $c / \sqrt{3}$ often used in calculations. We show here that if the peak in the radio light curve is associated with synchrotron self-absorption, this is simply not possible on energetic grounds (for three of our four sources, the power required for maximal size ejecta would be more than ten orders of magnitude greater than the Eddington limit). The reason for this discrepancy is that the assumed equipartition field for this maximal size is far below the field actually required to produce unity optical depth due to synchrotron self absorption in the source. We present instead an analysis which allows us to take into account this self-absorption condition, and to calculate the minimum energy for a source of unknown physical size.
We present sets of equations which allow the minimum energy, corresponding size, magnetic field and brightness temperature to be directly calculated from observables (Eqs. \ref{eqstart} -- \ref{eqend}). We furthermore provide approximate versions of the equations which can be used if only one frequency is available, but the optically thick condition is assumed (Eqs. \ref{eq1start} -- \ref{eq1end}). Although we do not explicitly consider the contribution from relativistic bulk motion for these minimisations, since it is extremely hard to estimate accurately for such jets, we demonstrate how it can be taken into account, and show that for nearly all parameters the estimates with no bulk relativistic motion remain lower limits. 

Taking this approach, we considered four well sampled optically thick outbursts from black hole X-ray binaries, covering a range of observing frequencies, luminosities and best-guess bulk Lorentz factors. 
Although it was not clear {\em a priori} that it would be the case, the minimum energy condition occurs within the allowed range of effective expansion speeds for all of the four BHXRB events considered here, $0.03 \leq \beta_e \leq 0.7$. In fact for three of the four events $\beta_e \leq 0.1$. This implies that the ejecta are constrained to expand much more slowly than the relativistic expansion speeds of $3^{-1/2} \leq \beta_e \leq 1$ which are often assumed, unless the filling factor is instead very small. If the expansion is indeed strongly constrained, the origin of this constraint it unclear; however we note that Tetarenko et al. (2017) also derived comparably low expansion speeds from detailed modelling of multifrequency events from V404 Cyg (and also discuss some possible origins of the effect). The minimum powers for energy injection into the synchrotron-emitting plasmas, typically $\sim 10^{36}$ erg s$^{-1}$, are easily reconciled with the overall power output for an accreting stellar mass black hole, particularly since such flare events typically occur close to the Eddington limit, which is $\sim 10^{39}$ erg s$^{-1}$ for a $\sim 7$M$_{\odot}$ black hole in a BHXRB. The real powers may of course be much greater.

Parameter estimates for optically thick flares based on the methods presented in this paper are fraught with major uncertainties. These include a lack of knowledge of the extent of the underlying electron spectrum, the nature and distribution of the positively charged component (positrons or protons) and, in most cases, a good measure of the bulk speed. It is alo highly unlikely that the events we observe as flares are truly associated with single, homogeneous component as envisaged in the van der Laan model. It is in fact made explicitly clear in e.g. Tetarenko et al. (2017) that events which look singular at cm wavelengths originate in multiple events at mm wavelengths. Nevertheless, considering the other uncertainties, and the clear smooth evolution from optically thick to optically thin observed for most such flares, it is a reasonable assumption that this evolution is representative of the majority of the emission. Therefore the approach here is a clear improvement on the naive assumption of highly relativistic expansion speeds, and allows a better estimate of the physical parameters, and minimum energy, of ejecta. 

Finally, we note that these calculations are applicable to any radio flare in which the peak is determined by synchrotron self-absorption and there the bulk motion is not highly relativistic. This may include radio flares from neutron star X-ray binaries (e.g. Migliari \& Fender 2006; Motta \& Fender 2019), cataclysmic variables (e.g. Coppejans et al. 2016; Mooley et al. 2017) and other related sources (e.g. Corbel et al. 2015) if and when they can be established to be optically thick. In order to build a large sample of events and to progress the field, many more observations of flaring events are required with enough spectral coverage to measure the spectral evolution. Fortunately most new radio telescope receivers are broad band and an in-band spectral index can often be measured. High cadence observations, in which multiple flares can be observed over a period of time, are also required in order to reliably estimate the time-averaged minimum power, and this may be harder to achieve on large facilities such as SKA.

\section*{Acknowledgements}
We would like to thank Alex Tetarneko for comments, and Lauren Rhodes for a thorough reading of an earlier version of this manuscript, which highlighted some errors in the text.

\newpage
\appendix
\section{Constants}

The constants $c_1$, $c_2$, $c_5$ and $c_6$ are from Pacholcyzk (1970) are are:

\smallskip
\noindent
$c_1 = \frac{3 e}{4 \pi m^3 c^5} = 6.27 \times 10^{18}$

\smallskip
\noindent
$c_2 = \frac{2 e^4}{3 m^4 c^7} = 2.37 \times 10^{-3}$

where $e$ is the charge on the electron, $m$ is the mass of the electron and $c$ is the speed of light.

The constants $c_5$ and $c_6$ are more complex functions, and are given numerically below for three different values of the electron energy index $p$.

\smallskip
\noindent
$c_5 = (2.26, 1.37, 0.97) \times 10^{-23}$ for $p=(1.5,2.0,2.5)$

\smallskip
\noindent
$c_6 = (9.69, 8.61, 8.10) \times 10^{-41}$ for $p=(1.5,2.0,2.5)$

\smallskip
\noindent

As usual, we consider a power-law distribution of synchrotron-emitting electrons of the form $N(E) dE \propto E^{-p} dE$, leading to the spectral index $\alpha = (1-p)/2$ when optically thin. 
The pseudo constant $c_{12}$ depends upon the upper and lower frequency bounds and slope of the observed synchrotron emission:

\begin{equation}
c_{12} = c_2^{-1} c_1^{1/2} \tilde{c}(p, \nu_1, \nu_2)$$
\end{equation}

where

\begin{equation}
\tilde{c}(p, \nu_1, \nu_2) = \frac{(p-3)}{(p-2)} \frac{\nu_1^{(2-p)/2}-\nu_2^{(2-p)/2}}{\nu_1^{(3-p)/2}-\nu_2^{(3-p)/2}}
\end{equation}

\noindent
Table A1 lists the values of $c_{12}$ calculated for each of the flares considered in this paper.

\begin{table}
\begin{tabular}{cc}
Flare & $c_{12}$ \\
\hline
1 & $1.1 \times 10^7$ \\
2 & $3.4 \times 10^6$\\
3 & $1.6 \times 10^7$\\
4 & $1.3 \times 10^7$\\
\hline
\end{tabular}
\caption{Values of the pseudo constant $c_{12}$ for each of the four flares considered in this work (see also Appendix 2, Table 8 of Pacholcyzk (1970).}
\end{table}

\section{Quantifying approximations}

Table \ref{lest} shows how the integrated luminosity between two frequencies for each of the flare events considered compares to the peak luminosity $\nu L_{\nu}$ which could be estimated from one frequency only.

\begin{table}
\begin{tabular}{cccc}
Flare & $L$ & $L_{\nu1}$ & $L_{\nu2}$ \\
\hline
1 & $9.6 \times 10^{31}$ & $1.4 \times 10^{31}$ & $9.7 \times 10^{31}$\\
2 & $6.9 \times 10^{33}$ & $9.6 \times 10^{33}$ & $1.7 \times 10^{34}$\\
3 & $6.3 \times 10^{33}$ & $1.6 \times 10^{33}$ & $5.7 \times 10^{33}$\\
4 & $1.0 \times 10^{31}$ & $7.0 \times 10^{30}$ & $1.9 \times 10^{31}$\\
\hline
\end{tabular}
\caption{Comparison of accurately calculated luminosity with specific luminosity $4 \pi D^2 \nu L_{\nu}$ using either the upper or lower observed frequency.}
\label{lest}
\end{table}

\begin{figure*}
\epsfig{file=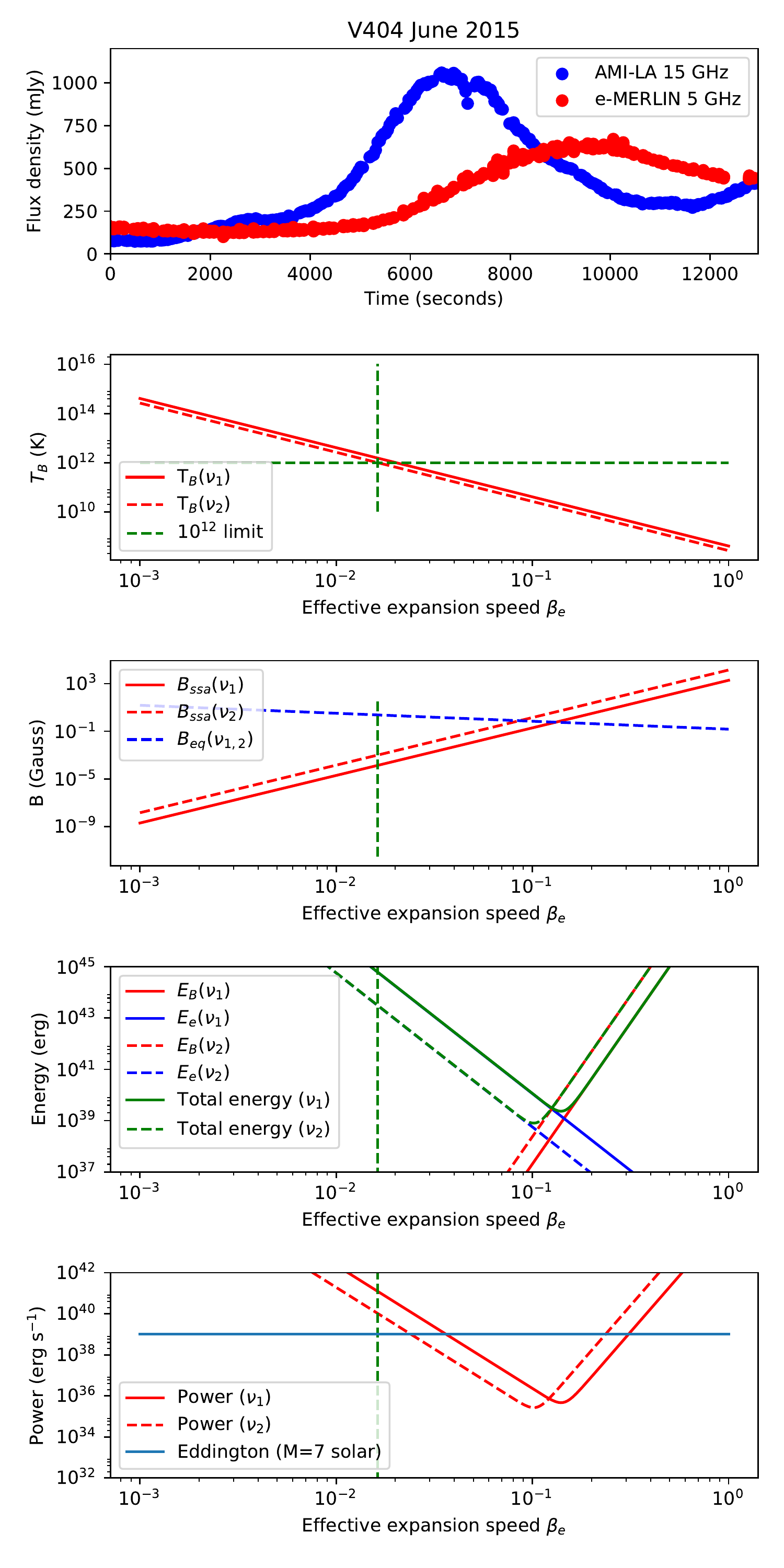, width=8cm}\quad\epsfig{file=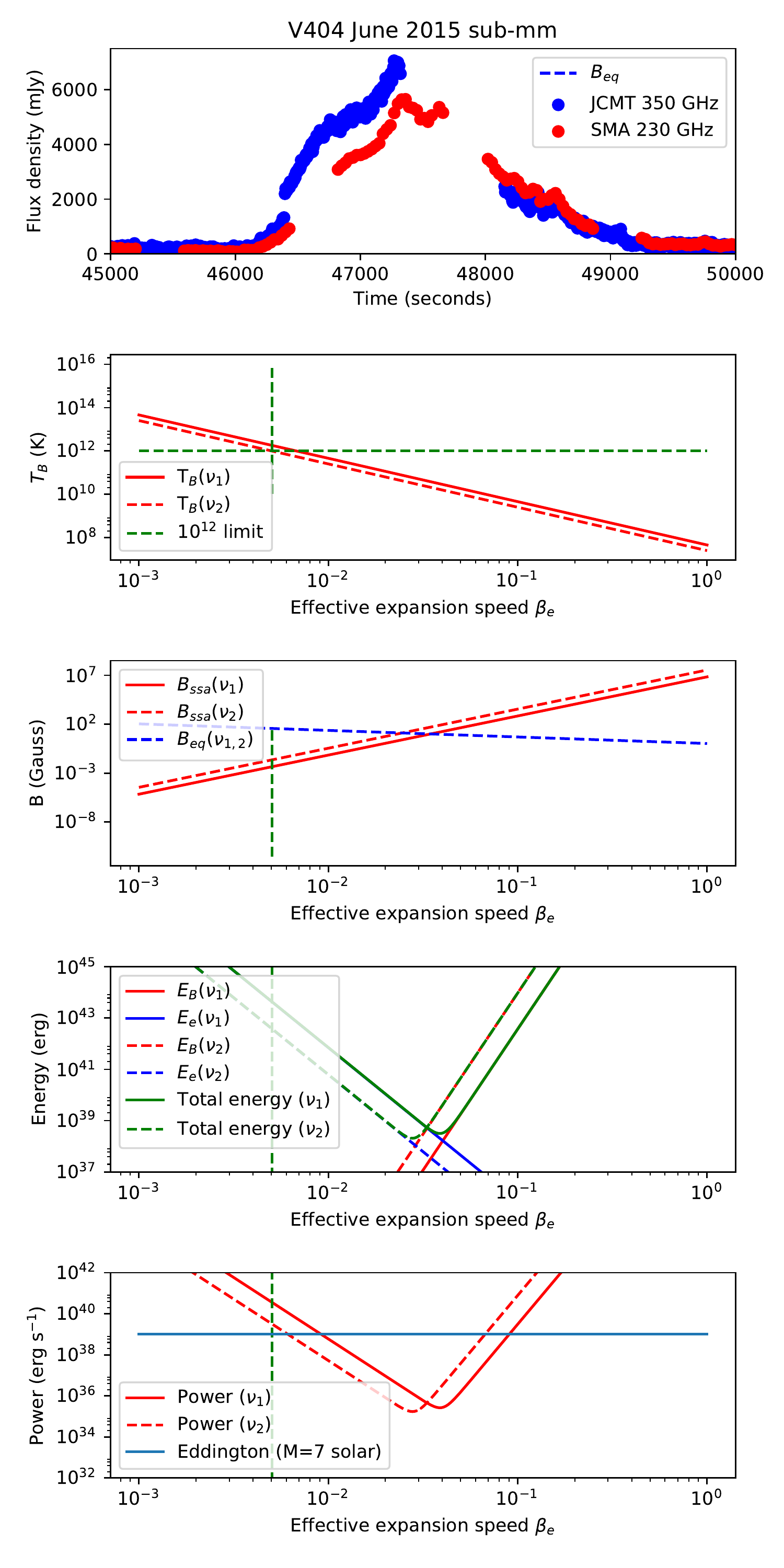, width=8cm}
\caption{As Fig 1 in main text, but with solutions for both the lower- and higher-frequency peaks plotted. The solutions are very similar for both sources.}
\end{figure*}

\setcounter{figure}{0}
\begin{figure*}
\epsfig{file=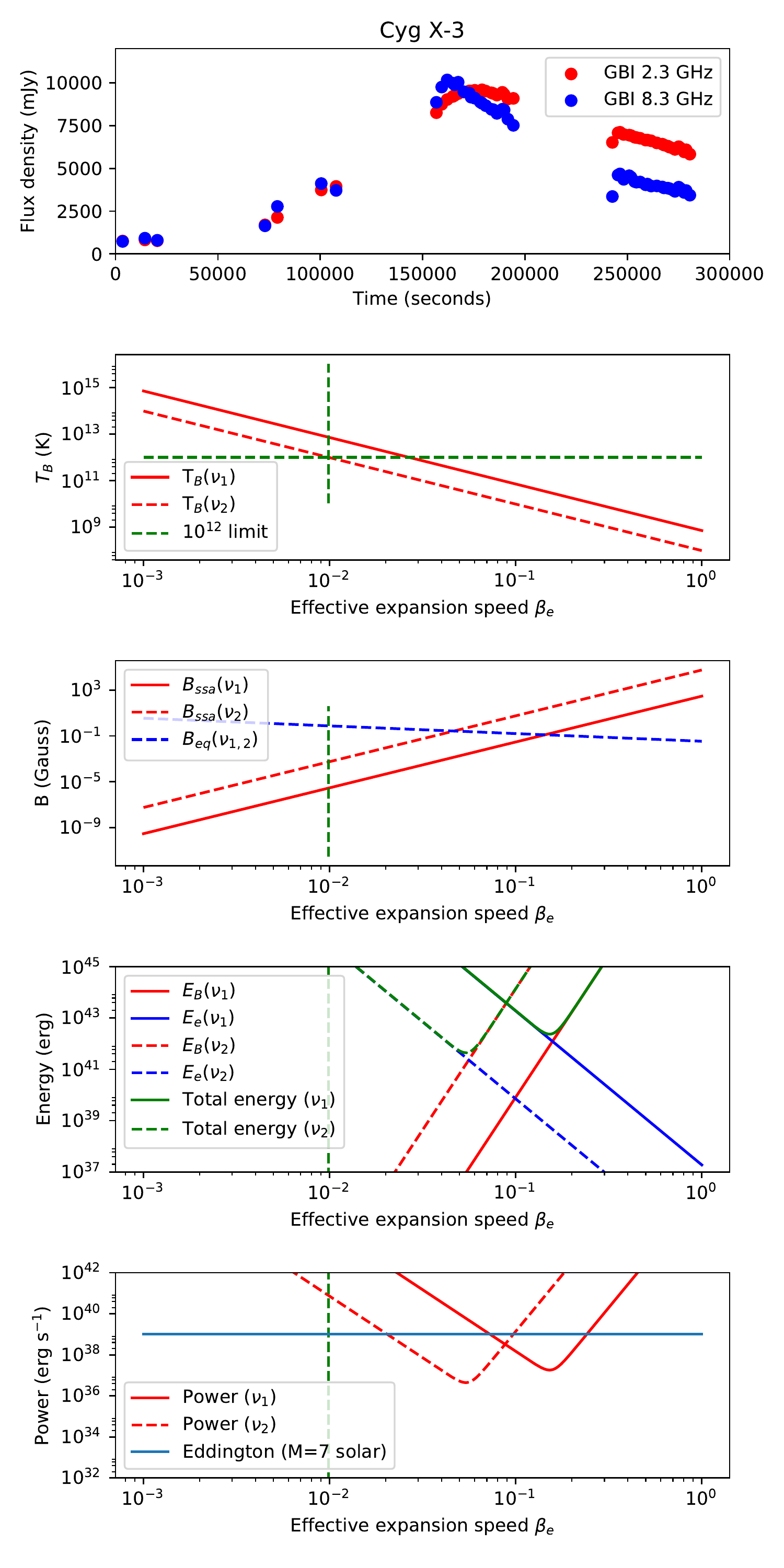, width=8cm}\quad\epsfig{file=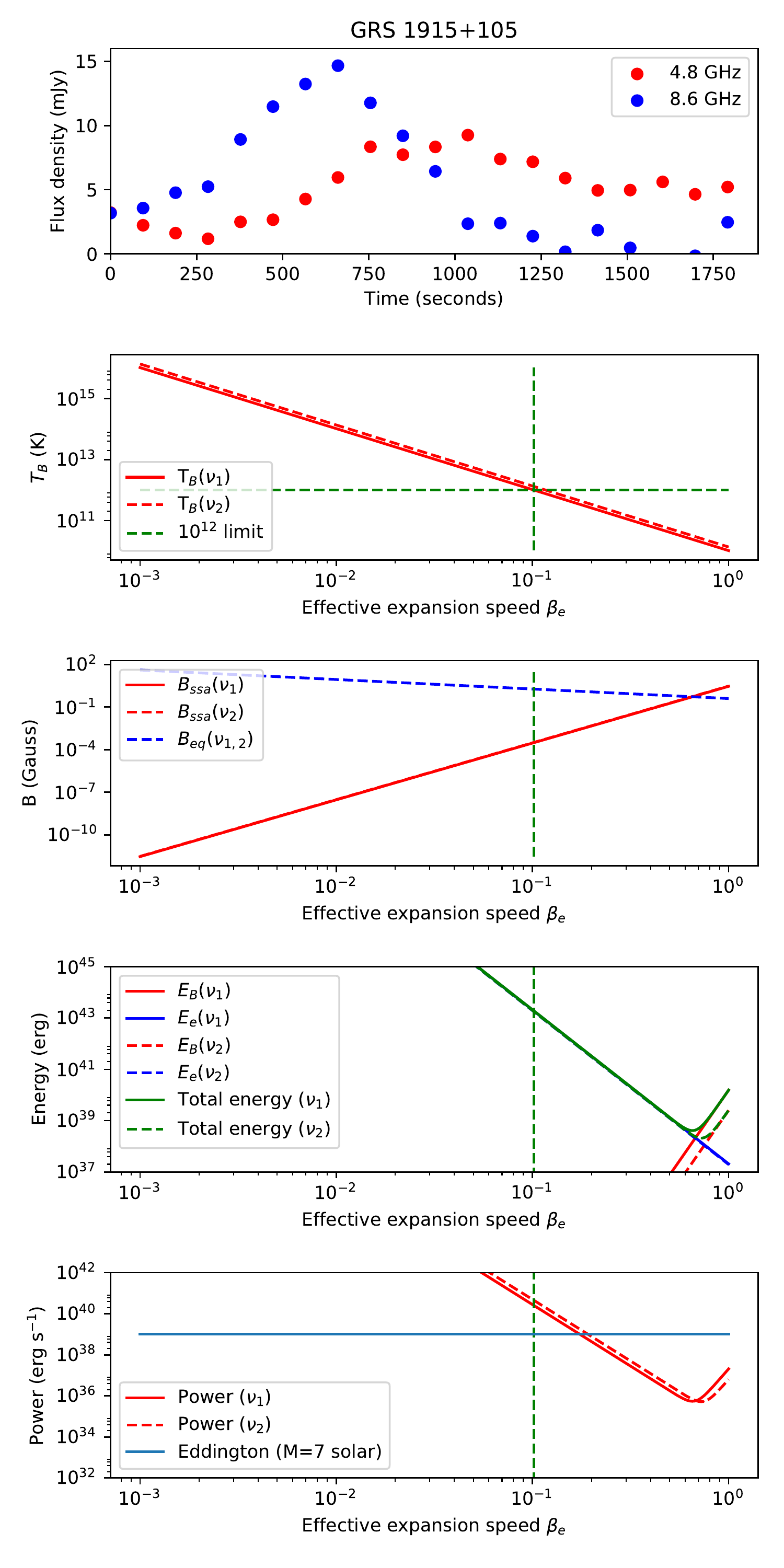, width=8cm}
\caption{{\bf (b)} As Fig 1 in main text, but with solutions for both the lower- and higher-frequency peaks plotted. The solutions are very similar for GRS 1915+105, but more significantly different for Cyg X-3.}
\end{figure*}

In Fig B.1 we plot the solutions for both frequencies, in the format of Fig 1. For three of the events (both V404 flares, and GRS 1915+105), the solutions are extremely similar (as can also be discerned from Table 1). For Cyg X-3 there is a significant difference in inferred expansion speed, although the other derived parameters are similar. The comparison of the two frequencies can provide a further indication of the uncertainty associated with single frequency estimates.

\bsp	
\label{lastpage}
\end{document}